\documentclass[10pt,conference,letterpaper]{IEEEtran}

\ifCLASSINFOpdf
 \usepackage[pdftex]{graphicx}
\else
   \usepackage[dvipdf]{graphicx}
   \DeclareGraphicsExtensions{.eps}
\fi
\usepackage{dblfloatfix}

\usepackage{epstopdf}

\newcommand{\argmax}{\arg\!\max}

\usepackage{epstopdf}
\usepackage{color}

\usepackage{graphicx}
\usepackage{caption}
\usepackage{subcaption}

\usepackage{amsmath, amsthm, amssymb} 
\usepackage{upgreek} 

\usepackage{algorithm} 
\usepackage{algpseudocode} 

\let\labelindent\relax
\usepackage{enumitem}
\floatstyle{plain}
\newfloat{myalgo}{tbhp}{mya}
{\begin{myalgo}[#1]
\centering
\begin{minipage}{#2}
\begin{algorithm}[H]}%
{\end{algorithm}
\end{minipage}
\end{myalgo}}
\usepackage{setspace}

\begin{document}
%
\title{Combined shared/dedicated resource allocation for Device-to-Device Communication}

\author{\IEEEauthorblockN{Pavel Mach, Zdenek Becvar}
\IEEEauthorblockA{Dpt. of Telecommunication Eng., Faculty of Electrical Engineering, 
Czech Technical University in Prague, Czech Republic\\
machp2@fel.cvut.cz, zdenek.becvar@fel.cvut.cz}
}


%


\maketitle

\begin{abstract}
Device-to-device (D2D) communication is an effective technology enhancing spectral efficiency and network throughput of contemporary cellular networks. Typically, the users exploiting D2D reuse the same radio resources as common cellular users (CUEs) that communicate through a base station. This mode is known as shared mode. Another option is to dedicate specific amount of resources exclusively for the D2D users in so-called a dedicated mode. In this paper, we propose novel combined share/dedicated resource allocation scheme enabling the D2D users to utilize the radio resources in both modes simultaneously. To that end, we propose a graph theory-based framework for efficient resource allocation. Within this framework, neighborhood relations between the cellular users and the D2D users and between the individual D2D users are derived to form graphs. Then, the graphs are decomposed into subgraphs to identify resources, which can be reused by other users so that capacity of the D2D users is maximized. The results show that the sum D2D capacity is increased from 1.67 and 2.5 times (depending on a density of D2D users) when compared to schemes selecting only between shared or dedicated modes.

\end{abstract}

\begin{IEEEkeywords}
device-to-device communcation, mode selection, resource allocation, shared allocation, dedicated allocation.
\end{IEEEkeywords}

\IEEEpeerreviewmaketitle

\section{Introduction}
Device-to-device (D2D) communication is an emerging paradigm enhancing spectrum and energy efficiency of mobile communications systems [1]. The D2D enables direct communication of two devices in proximity without involvement of a base station, which is typically denoted as eNB in mobile networks. The direct communication allows saving radio resources by avoiding two-hop transmission of the data (to eNB and to the user). The D2D can be classified according to spectrum utilization into an in-band D2D (using license radio resources allocated to conventional cellular users) or an out-band D2D (exploiting unlicensed frequency bands, such as WiFi direct, ZigBee or Bluetooth) [2].

The D2D users (DUEs) accessing license band (i.e., in-band D2D) can use three communications modes: 1) $cellular$ mode, 2) $dedicated$ mode, also known as overlay, and 3) $shared$ mode,  also known as underlay. In case of the cellular mode, the DUEs communicate through the eNB as in a conventional cellular network. The dedicated mode is distinguished by the fact that the DUEs access dedicated resources with respect to the cellular users (CUEs). Consequently, interference between the DUEs and the CUEs is efficiently avoided, but the system may experience a low spectral efficiency or an insufficient amount of resources for both types of users. In case of the shared mode, the DUEs reuse the same resources as the CUEs. Although this mode enables the highest frequency reuse when compared to other two modes, interference between the CUEs and the DUEs is the most challenging aspect here. 

In recent years, significant effort has been invested to address the problem of mode selection, that is, the selection of the most profitable mode for the DUEs and efficient resources allocation. The research works related to the mode selection can be classified into papers selecting: 1) between the cellular mode and the shared or dedicated mode [3]-[8], 2) between the dedicated and shared modes [9][10], and 3) among the cellular, dedicated, and shared modes [11]-[13]. Most of the existing works focusing on the mode selection and resource allocation for DUEs expect that each DUE uses just one communication mode (cellular, dedicated, or shared). However, in [7], the authors propose to allow the DUEs to access the radio resources in the cellular and shared modes at the same time. The paper addresses routing problem deciding whether data should be routed through eNB (cellular mode) or transmitted directly between DUEs using shared mode. 

In this paper, we propose a Combined Shared/Dedicated resource allocation scheme labeled as CSD. The key difference between [7] and CSD is that instead of routing problem we address resource allocation problem when the DUEs always communicate directly accessing resources in both shared and dedicated mode simultaneously. To that end, the main objective of the proposed CSD is to maximize the capacity of DUEs, while the performance of the CUEs is not impaired due to power restriction of the DUEs in the shared mode. To increase the resource utilization by the DUEs, several DUEs can access the same resources if they are not neighbors, that is, if interference among them is below a predefined threshold ($\tau_N$). In this regard, we propose a novel graph theory-based framework for determination of two neighborhood relations: i) between the CUEs and the DUEs, and ii) among the DUEs. The neighborhood relations are then exploited for the allocation of resources to the D2D pairs. 

The rest of the paper is organized as follows. The next section describes the system model. The graph theory based framework for the proposed resource allocation is introduced in Section III. The Section IV explains the proposed resource allocation. Section V is dedicated for description of the simulation scenario and a discussion of the simulation results. The last section gives our conclusion and future works.

\section{System model}
This section describes system model. We assume a single cell scenario with one eNB. Within coverage area of the eNB, $C$ CUEs and $D$ D2D pairs exploiting uplink cellular network resources are deployed. Each D2D pair is composed of one DUE transmitter (DUE-T) sending data to the DUE receiver (DUE-R). The DUEs of the same D2D pair always communicate directly, i.e., the cellular  is not applied for the DUEs. The allocation of resources to the DUEs is fully controlled by the eNB as considered, e.g., in [11][13]. 	

We assume general multiple access technology, such as OFDMA or SC-FDMA, where the available bandwidth is divided into $n$ channels, represented in our case by resource blocks (RBs). The whole bandwidth is split into shared and dedicated regions. The shared region contains $n_s$ RBs that are accessible by both the CUEs and DUEs. Consequently, interference between the CUEs and the DUEs has to be taken into consideration in this region. In contrast, the dedicated region is composed of $n_d$ RBs, which are accessible only by the DUEs and, thus, there is no mutual interference between the DUEs and the CUEs.

The SINR observed by the eNB at the $r$-th RB is calculated as:
\begin{eqnarray}
\gamma_{e}^r=\frac{{g_{Ce}^r}{P_{C}^r}}{NI^r+\sum_{
i=1}^{D^r}g_{T_ie}^r{P_{i}^{s,r}}}
\label{eqn:3}
\end{eqnarray}
where $g_{Ce}^r$ is the channel gain between the CUE and eNB at the $r$-th RB, $P_C^r$ represents the transmission power of the CUE at the $r$-th RB, $NI^r$ stands for the thermal noise plus interference observed by the eNB from adjacent cells at the $r$-th RB, $g_{T_ie}^r$ is the channel gain between the $i$-th DUE-T and the eNB at the $r$-th RB, $P_i^{s,r}$ corresponds to the transmission power of the $i$-th DUE-T in the shared region at the $r$-th RB, and $D^r$ is the number of DUE-Ts transmitting at the $r$-th RB, because the proposed scheme allows to reuse the same RB by more than one D2D pair if interference between the D2D pairs is below a predefined threshold as explained later.

In the shared region, the transmitting power of the DUE-T is restricted in order to limit interference caused to the CUEs. The restriction of the transmission power for each DUE-T in the shared region is defined by the eNB so that the signal received by the eNB from the i-th DUE-T ($RSS_{di}$) is:
\begin{eqnarray}
RSS_{di}=\frac{NI_e}{\tau_{DUE}}
\label{eqn:3}
\end{eqnarray}
where $\tau_{DUE}$ is the D2D threshold defining the amount of interference caused by the DUE-T with respect to the noise plus interference from the adjacent eNBs measured at the given eNB ($NI_e$). Since we assume $\tau_{DUE}=10$, the level of interfering signal from the DUE-T received at the eNB is 10 times smaller than the level of interference from other sources. Hence, the performance of the CUEs can be considered as unimpaired at all by the proposed algorithm. Then, SINR observed by the DUE-R of the $j$-th D2D pair at the $r$-th RB in the shared region can be expressed as:
\begin{eqnarray}
\gamma_{j}^{s,r}=\frac{{g_{T_jR_j}^r}{P_{j}^{s,r}}}{NI^r+\sum_{\substack{
i=1 \\ i \neq j}}^{D^r}g_{T_iR_j}^r{P_{i}^{s,r}}+g_{CR_j}^rP_C^r}
\label{eqn:3}
\end{eqnarray}
where $g_{T_jR_j}^r$ is the channel gain between the DUE-T and the DUE-R of the same $j$-th D2D pair at the $r$-th RB, $g_{T_iR_j}^r$ stands for the channel gain between the $i$-th DUE-T and the $j$-th DUE-R at the $r$-th RB, $g_{CR_j}^r$ represents the channel gain between the CUE and the $j$-th DUE-R at the $r$-th RB, $P_j^{s,r}$ and $P_i^{s,r}$  are the transmission powers of the $j$-th and $i$-th DUE-T at the $r$-th RB, respectively. From (3), we can observe that interference to the D2D pairs in the shared region originates from other D2D pairs reusing the same resources (similarly as in (1)) and by the CUEs, which occupy the reused resources.

The SINR observed by the $j$-th DUE-R at the $r$-th RB in the dedicated region is defined as:
\begin{eqnarray}
\gamma_{j}^{d,r}=\frac{{g_{T_jR_j}^r}{P_{j}^{d,r}}}{NI^r+\sum_{\substack{
i=1 \\ i \neq j}}^{D^r}g_{T_iR_j}^r{P_{i}^{d,r}}}
\label{eqn:3}
\end{eqnarray}
where $P_j^{d,r}$ is the transmission power of the $j$-th DUE-T in the dedicated region and $P_i^{d,r}$ is the transmission power of the $i$-th DUE-T, which causes interference to the $j$-th D2D pair. The transmission power of the DUE-Ts in the dedicated region is not restricted as in the shared region, because the dedicated resources are not shared with the CUEs. Note that in the simulation, we consider several $P_i^{d,r}$ values and we analyze its impact on the performance.

\section{Graph theory-based framework for proposed CSD}	
To allocate resources to the D2D pairs properly by the proposed CSD and to avoid harmful interference, the eNB has to be aware of: 1) the list of D2D pairs that can reuse RBs already assigned to the CUEs, and 2) the list of RBs that can be reused by more than one D2D pair. Both above-mentioned aspects are determined by the eNB through the knowledge of two types of the neighborhood relations: 1) between the CUEs and the DUEs, and 1) among the DUEs. The following subsections describe a determination of the CUEs’ and the DUEs’ neighbors and explain exploitation of the neighborhood relations for the resource allocation to the D2D pairs. 

\subsection{Determination of CUE and DUE neighbors}
The DUEs may reuse RBs allocated to the CUEs that are not their neighbors. In the proposed scheme, the CUE is considered to be a neighbor of the $j$-th D2D pair if:
\begin{eqnarray}
\gamma_{j}^{s,r}<\gamma_{min}
\label{eqn:3}
\end{eqnarray}
where $\gamma_{min}$ is minimal SINR guaranteeing reliable communication. Notice that the classification whether the CUE is neighbor or not does not solely depends on signal received from the CUE at the DUE-R. Even distant CUE can be considered as the neighbor just because the received signal quality is low. This situation can happen, for example, if the D2D-T and the D2D-R are far away from each other or if the D2D-T is in vicinity of the eNB and its transmission power is restricted according to (2) to avoid interference to the eNB. Moreover, the CUE is considered to be a neighbor even if DUE-R is strongly interfered from the CUEs belonging to other cells since the D2D pair is not able to reuse these resources due to interference.

The $\gamma_j^{s,r}$ is obtained by the eNB from channel quality report, e.g., by means of CSI reports as defined in LTE [14], sent by individual DUE-R. In this regard, the eNB dedicates specific intervals when the D2D-R should listen to the CUEs' transmissions and determine which RBs can be reused in the shared region. During these intervals, the D2D-T cannot use the RBs in the shared region for data transmission. Nevertheless, the effect on throughput is negligible as these intervals can be scheduled only sparingly (e.g., one interval with duration of 1 ms is dedicated for this purpose per one or several seconds depending on speed of the CUEs).

Besides the CUEs' neighbors, each D2D pairs may have several DUE neighbors. The D2D pairs may reuse the same RBs if they are not mutual neighbors, that is, if interference among them is below the predefined threshold. In this paper, the $i$-th D2D pair is considered to be the neighbor of the $j$-th D2D pair if the received signal strength from the $i$-th DUE-T at the side of the $j$-th DUE-R is: 
\begin{eqnarray}
RSS_{ij}>\frac{NI_{R_j}}{\tau_{N}}
\label{eqn:3}
\end{eqnarray}
where $NI_{R_j}$ is the thermal noise plus interference observe at the $j$-th DUE-R and $\tau_N$ is the threshold distinguishing neighbors both in the shared and dedicated regions. The threshold $\tau_N$ allows to adjust the level of interference among the D2D pairs and the amount of D2D pairs, which are able to reuse the same RBs. If $\tau_N<1$, the $i$-th D2D pair is the neighbor of the  $j$-th pair despite the fact that it generates higher interference than $NI_j$ resulting in a higher reuse of the RBs. If $\tau_N>1$, interference among the D2D pairs is lower than $NI_j$, but the reuse factor is decreased as well. The proper selection of $\tau_N$ could be seen as an optimization problem. Hence, if we define $C^s$ and $C^d$ as capacities in the shared and dedicated regions, the objective is to find $\tau_{N,opt}$ so that:
\begin{eqnarray}
[\tau_{opt}]=\argmax_{\tau_N}({C^s+C^d}), {\tau_N\in\{\tau_{N,min},\tau_{N,max}\}}
\label{eqn:3}
\end{eqnarray}
Due to the limited space, we leave a derivation of $\tau_{N,opt}$ analytically for future research and we just investigate its impact on the performance by simulations.

The D2D pairs can find their neighbors by means of discovery procedure proposed in [15], where the DUE-R measures experienced SINR of the received discovery messages send by other DUE-Ts. Then, the D2D-R sends SINR report to the eNB. Notice that the DUE-R may send reports from measurement of the CUE and the DUE neighbors at the same time.

\subsection{Forming graphs, graph decomposition into subgraphs, and determination of maximal cliques}
Based on the channel quality reported by the DUE-Rs for determination of the neighborhood relations, as explained in previous subsections, two graphs are created: one for shared region, one for the dedicated region. The graph for shared region is denoted as $G^s=(V_x^s,E_x^s)$, where $V_x^s$ represents individual D2D pairs that are able to reuse the RBs in the shared region (vertices of graph), and $E_x^s$ represents interference between the D2D pairs in the shared region. Analogously, the graph for dedicated region is denoted as $G^d=(V_x^d,E_x^d)$, where $V_x^d$ represents the D2D pairs using the RBs in the dedicated region and $E_x^d$ shows neighborhood relations (by means of interference) between the D2D pairs in the dedicated region.
 
To allow the eNB allocate the RBs to the D2D pairs in both regions, the eNB further decomposes $G^s$ and $G^d$ into a set of subgraphs so that each subgraph contains only D2D pairs that can potentially reuse the same RBs. 
In case of $G^s$, the individual subgraphs ($G^{s,1}, G^{s,2},..., G^{s,C}$) are composed of the D2D pairs that are able to reuse the RBs assigned to the same CUE. Consequently, the number of subgraphs in shared region is exactly the same as the number of the CUEs. Moreover, the number of RBs available for each subgraph corresponds to the number of RBs assigned to individual CUEs. 
In case of $G^d$, the individual subgraphs ($G^{d,1}, G^{d,2},..., G^{d,D}$) are composed of the D2D pairs that can reuse the RBs assigned to other D2D pairs in the dedicated region (i.e., one subgraph is created for each D2D pair). Notice that in the dedicated region, the D2D pairs receives always some RBs allocated to them by default and the RBs that can be reused from other D2D pairs. 

To properly allocate the RBs to the D2D pairs within each subgraph, the maximal cliques in all subgraphs are found by the eNB. In general, the clique in $G$ is defined as a subset of vertices (in our case subset of D2D pairs), $C \subseteq V$, such that resulting subgraph is a complete graph. Then, the clique is called a maximal clique in $G$ if there is no clique $C'$ such that $C' \supset C$. In other words, the maximal clique is not a subset of any other cliques in $G$. As a consequence, the D2D pairs in the same maximal clique cannot reuse the same RBs. The eNB finds all maximal cliques in $G^s$ and $G^d$ subgraphs by means of Bron-Kerbosch algorithm [16]. The complexity of the algorithm is in the worst case $O(3n/3)$, where n is the number of D2D pairs. Despite the fact that Bron-Kerbosch algorithm is NP-hard, it can be used even in large networks as demonstrated in [17].

\begin{figure}[b!]
\centering
\begin{subfigure}[t]{0.4\textwidth}
\centering
\includegraphics[width=\textwidth]{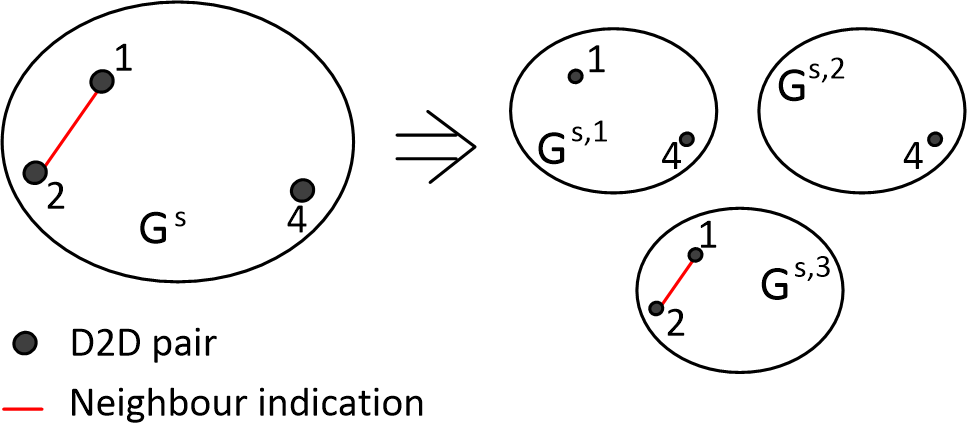}
\caption{}
\end{subfigure}
\\
\begin{subfigure}[t]{0.4\textwidth}
\centering
\includegraphics[width=\textwidth]{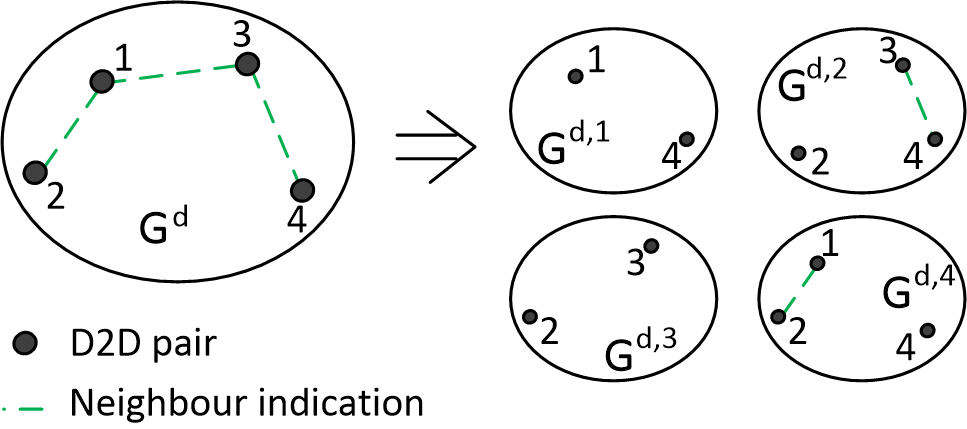}
\caption{}
\end{subfigure}
\caption{Example of $G^s$ and $G^d$ determination and their decomposition into subgraphs and maximal cliques.}
\label{fig:1}
\end{figure}

The creation of the graphs $G^s$ and $G^d$ and their further decomposition into subgraphs and determination of the  maximal cliques is illustrated in Fig.~\ref{fig:1}. In Fig.~\ref{fig:1}a, $G^s$ is composed of the D2D pair 1, 2 and 4 as these can utilize the RBs in the shared region according to (5). Then, three subgraphs are created (each for single CUE): $G^{s,1}$ indicating that the D2D pair 1 and 4 can reuse the RBs allocated to the CUE 1, $G^{s,2}$ saying that resources used by the CUE 2 can be exploited by the D2D pair 4, and $G^{s,3}$ showing that the D2D pair 1 and 2 can reuse the RBs allocated to the CUE 3. In Fig.~\ref{fig:1}a,  $G^d$ is decomposed into four subgraphs (each for one D2D pair): $G^{d,1}$ indicating that the RBs assigned to the D2D pair 1 can be fully reused by the D2D pair 4, $G^{d,2}$ showing the RBs for the D2D pair 2 can be reused by the D2D pair 3 and 4, etc. Finally, the eNB finds all maximal cliques within each subgraph ($N_{mc}$). For example, there are two maximal cliques $\{1\}$ and $\{4\}$ in $G^{s,1}$ ($N_{mc}^{s,1}=2$), one maximal clique $\{4\}$ for $G^{s,2}$ ($N_{mc}^{s,2}=1$), and one maximal clique $\{1,2\}$ for $G^{s,3}$ ($N_{mc}^{s,3}=1$). Analogously, there are two maximal cliques $\{1\}$ and $\{4\}$ in $G^{d,1}$ ($N_{mc}^{d,1}=2$), two maximal cliques $\{2\}$ and $\{3,4\}$) exists in $G^{d,2}$ ($N_{mc}^{d,2}=2$), etc.

\subsection{Formulation of objectives for the proposed CSD scheme}
The objective of the proposed CSD resource allocation scheme is to maximize sum D2D capacity ($C$). The $C$ is a sum of the capacities in the shared region ($C^s$) and in the dedicated region ($C^d$). As the $C^s$ and $C^d$ are independent, we can maximize these two capacities separately. Hence, the first objective is to maximize $C^s$ as:
\begin{eqnarray}
\begin{aligned}
C^s=\max\left(\sum_{z=1}^C\left(\sum_{k=1}^{N_{mc}^{s,z}}\left(\sum_{r=1}^{n_{mc}^{s,z}}\Gamma_{k}^{s,z,r}\right)\right)\right) 
\\
\text{s.t.~~~~$n_{mc}^{s,z} \leq n_c^z$,~ $\forall z$~~~~~~~~~~~~~~~~~} \\
\text{$P_j^{s,r}$ restricted acc. (2)~~~~~~~~}
\label{eqn:3}
\end{aligned}
\end{eqnarray}
where $n_{mc}^{s,z}$ is the number of RBs in individual maximal clique, $N_{mc}^{s,z}$ represents the number of maximal cliques found in the individual subgraphs, $\Gamma_k^{s,z,r}$ is the transmission efficiency representing the number of bits transmitted in the $r$-th RB of the $k$-th maximal clique of the D2D pairs reusing resources assigned to the $i$-th CUE, and $n_c^z$ is the amount of RBs allocated to the $z$-th CUE. Note that $\Gamma_k^{s,z,r}$ is derived from $\gamma_j^{s,r}$ in the shared region. The capacity in shared region is restricted by the number of RBs available in each maximal clique ($n_{mc}^{s,z})$ and by the power constrains according to (2).

The second objective is to maximize $C^d$ as:
\begin{eqnarray}
\begin{aligned}
C^d=\max\left(\sum_{z=1}^D\left(\sum_{k=1}^{N_{mc}^{d,z}}\left(\sum_{r=1}^{n_{mc}^{d,z}}\Gamma_{k}^{d,z,r}\right)\right)\right) 
\\
\text{s.t.~~~~$n_{mc}^{d,z} \leq n_d^z$~ $\forall z$~~~~~~~~~~~~~~~~~~} 
\label{eqn:3}
\end{aligned}
\end{eqnarray}
where $N_{mc}^{d,z}$ is the number of maximal cliques found in individual subgraphs, $\Gamma_k^{d,z,r}$ stands for the transmission efficiency in the dedicated region, and $n_d^z$ is the amount of RBs allocated by default to the $z$-th D2D pair. The capacity in the dedicated region is restricted by the amount of RBs available in each maximal clique ($n_{mc}^{d,z}$) analogously to the shared region.

\begin{figure*}[t!]
\centering
\includegraphics[width=\textwidth]{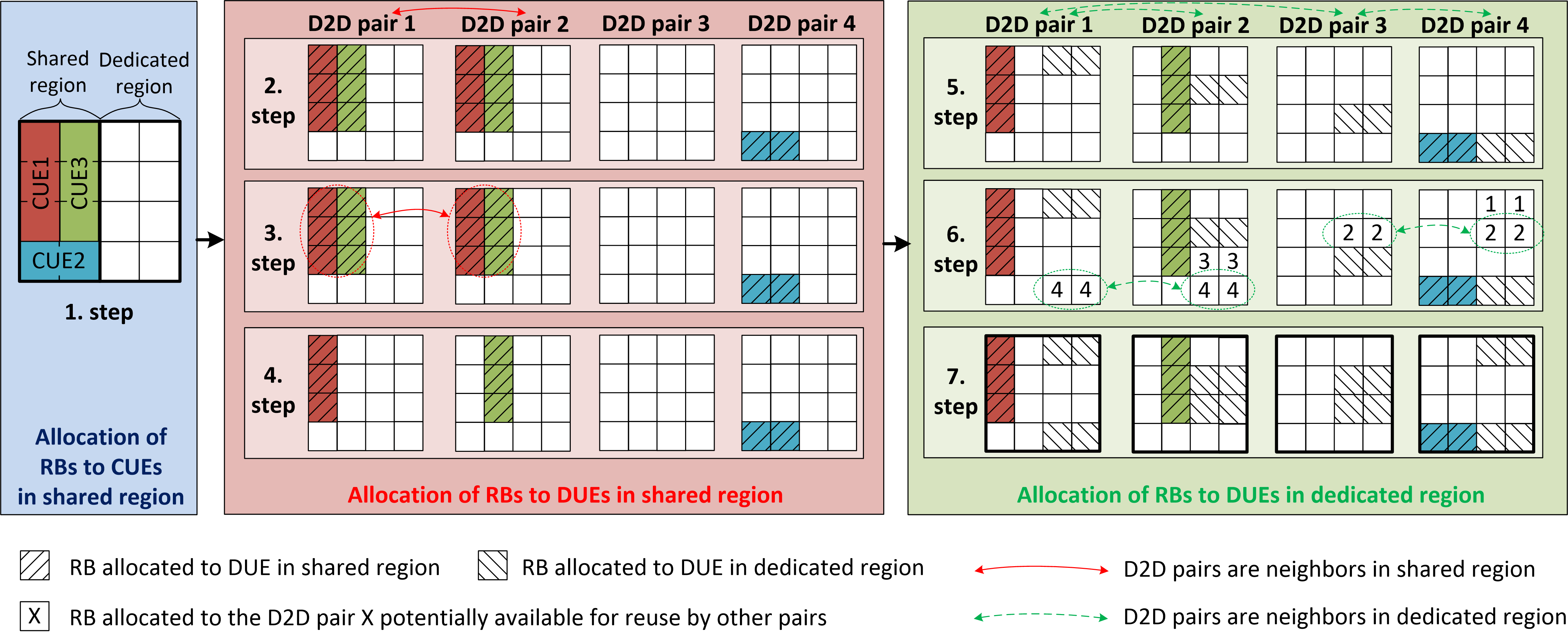}
\caption{Example of allocation process according to proposed allocation algorithm.}
\label{fig:2}
\end{figure*} 

\section{Algorithm for proposed CSD scheme}
The allocation of resources according to CSD scheme is composed of: 1) allocation of RBs to the CUEs as this directly impacts the amount of RBs in both shared and dedicated regions, 2) allocation of RBs to the DUEs in the shared region, and 3) allocation of the RBs available for the DUEs in the dedicated region. To this end, we propose the algorithm for allocation of resources that can be divided into seven subsequent steps:
\begin{enumerate}[leftmargin=0.45cm]
\item 
Allocation of $n_s$ RBs to the CUEs in the shared region, i.e., the region where RBs are accessible by both the CUEs and the DUEs. The amount of RBs in the shared region depends on two factors: 1) current requirements of the CUEs and 2) the amount of RBs that can be used by the DUEs in the shared and dedicated regions. The second factor takes into account the fact that not all D2D pairs may be able to reuse the RBs in the shared region due to the power restriction according to (2). For example, if the eNB would allocate all the RBs to the CUEs, some D2D pairs may not be able to communicate at all. In this paper we consider the same amount of RBs available in both shared and dedicated regions. A dynamic allocation of the amount of RBs in individual regions considering two above-mentioned factors is left for future research.
\item
Determination if individual D2D pairs can reuse the RBs allocated to the CUE(s) in the shared region if (5) is fulfilled (as explained in Section 3).
\item
Determination if multiple D2D pairs can reuse the same RBs assigned to the same CUE. This is done by finding all maximal cliques in individual subgraphs $G^{s,z}$, where $1\leq z\leq C$ (see Fig.~\ref{fig:1}a).
\item
Allocation of the resources to individual D2D pairs in the shared region maximizing $C^s$. This is done by allocation of the RBs to the D2D pairs with the highest transmission efficiency according to (8).
\item
Allocation of the default amount of RBs ($n_d^i$) in the dedicated region to each D2D pair, where the default number of  the RBs allocated to the $i$-th D2D pair $n_d^i$ is calculated as:
 \begin{eqnarray}
n_d^i=n_d\frac{N_{mc}^{d,i}}{\sum_{z=1}^{D}N_{mc}^{d,z}}
\label{eqn:3}
\end{eqnarray}
The $n_d^i$ is proportional to the amount of maximal cliques as this determines how many times the RBs allocated by default to the D2D pair can be reused in the dedicated region by the other D2D pairs. 
\item
Determination of the RBs, which can be potentially reused and by whom these can be reused depending on the allocation of RBs in the previous step and the knowledge of neighborhood relations among D2D pairs obtained from subgraphs $G^{d,z}$, where $1 \leq z \leq D$ (see Fig.~\ref{fig:1}b).
\item
Allocation of resources to individual D2D pairs in the dedicated region maximizing $C^d$. This is done by the allocation of the RBs to the D2D pairs with the highest transmission efficiency according to (9).
 \end{enumerate} 
 
The example of allocation process according to the proposed algorithm is shown in Fig.~\ref{fig:2}, where $n_s$ = $n_d$ = 8 RBs and neighborhood relations are taken from Fig.~\ref{fig:1}. The allocation process is as follows:
\begin{enumerate}[leftmargin=0.45cm]
\item 
The eNB allocates RBs to the CUEs in the shared region.
\item
The eNB identifies that the D2D pair 1 and 2 can reuse RBs allocated to the CUE 1 and CUE 3, while the D2D pair 4 reuses RBs assigned to the CUE 2. The D2D pair 3 is not able to reuse any RBs in the shared region.
\item
The eNB determines that the D2D pair 1 and 2 are mutual neighbors (i.e., orthogonal RBs have to be allocated to them in the shared region), while the D2D pair 4 has no neighbors that are able to reuse the RBs of the CUE 2. 
\item
According to (8), the eNB allocates all RBs assigned to the CUE 1 to the D2D pair 1 since the D2D pair 1 has higher transmission efficiency at these RBs than the D2D pair 2. Contrary, the D2D pair 2 reuses all RBs allocated to the CUE 3 as it experiences higher transmission efficiency at these RBs than the D2D pair 1. Finally, the D2D pair 4 reuses all RBs allocated to the CUE 2.
\item
The eNB allocates $n_d^i$ RBs to individual D2D pairs taken into account (10). According to Fig.~\ref{fig:1}b, all subgraphs contain two maximal cliques. Hence, each D2D pair obtains $n_d^i$ = 2 RBs by default.
\item
The eNB determines that the D2D pair 1 can reuse RBs allocated to the D2D pair 4, the D2D pair 2 can reuse the RBs assigned to the D2D pair 3 and 4, the D2D pair 3 can reuse the RBs allocated to the D2D pair 2, and the D2D pair 4 can reuse the RBs assigned to the D2D pair 1 and 2.
\item 
To meet (9), the eNB decides that the RBs allocated by default to the D2D pair 1 can be fully reused by the D2D pair 4 since the D2D pair 1 experiences higher transmission efficiency than the D2D pair 2 at these RBs. Similarly, the RBs allocated to the D2D pair 2 should be reused solely by the D2D pair 3 since it has higher transmission efficiency at these RBs than the D2D pair 4. Then, the RBs allocated by default to the D2D pair 3 are available for reuse by the D2D pair 2, and the RBs of the D2D pair 4 are reused by the D2D pair 1.
\end{enumerate}
 
\section{Simulations}
This section describes simulation scenario and models used for evaluation and, then, simulation results are presented and discussed.

\subsection{Simulation scenario}
The performance of the proposed scheme is evaluated in MATLAB simulator. We assume square simulation area with size of 500 m. The simulation area contains one eNB deployed in the middle of the area, 20 CUEs and up to 75 D2D pairs. Both the positions of the CUEs and the D2D pairs are generated randomly with uniform distribution. The maximum distance of the UEs creating D2D pair is set to 200 m. We assume that D2D pairs always communicate directly as communication through the eNB (i.e., cellular mode) introduces no benefits.
\begin{table}[t!]
	\footnotesize
	\caption{Parameters and settings for simulations}
	\label{tab:Tab1}
	\centering
	\begin{tabular}{|p{5.8cm}||p{2.1cm}| } 
		\hline
		{\bf Parameter} & {\bf Value} \\
		\hline
		Simulation area & 500x500\\
		\hline
		Number of CUEs & 20\\
		\hline
		Number of D2D pairs & 5--75\\
		\hline
		Max. distance between DUE-T and DUE-R & 200 m\\
		\hline
		Carrier frequency & 2 GHz\\
		\hline
		Channel bandwidth & 20 MHz \\
		\hline
		Signaling overhead & 25 \%\\
		\hline
		Number of RBs in shared/dedicated regions & 750/750\\
		\hline
		Transmission power of CUE and DUEs in dedicated region & 10, 15, 20 dBm\\
		\hline
		$\tau_N$ & -30--0\\
		\hline
		$\gamma_{min}$ & -9.478 dB\\
		\hline
		$\tau_{DUE}$ & 10 dBm\\
		\hline
		Number of simulation drops & 200\\
		\hline
        
	\end{tabular}
\label{tab}
\end{table}
\begin{figure}[b!]
\centering
\includegraphics[scale=0.43]{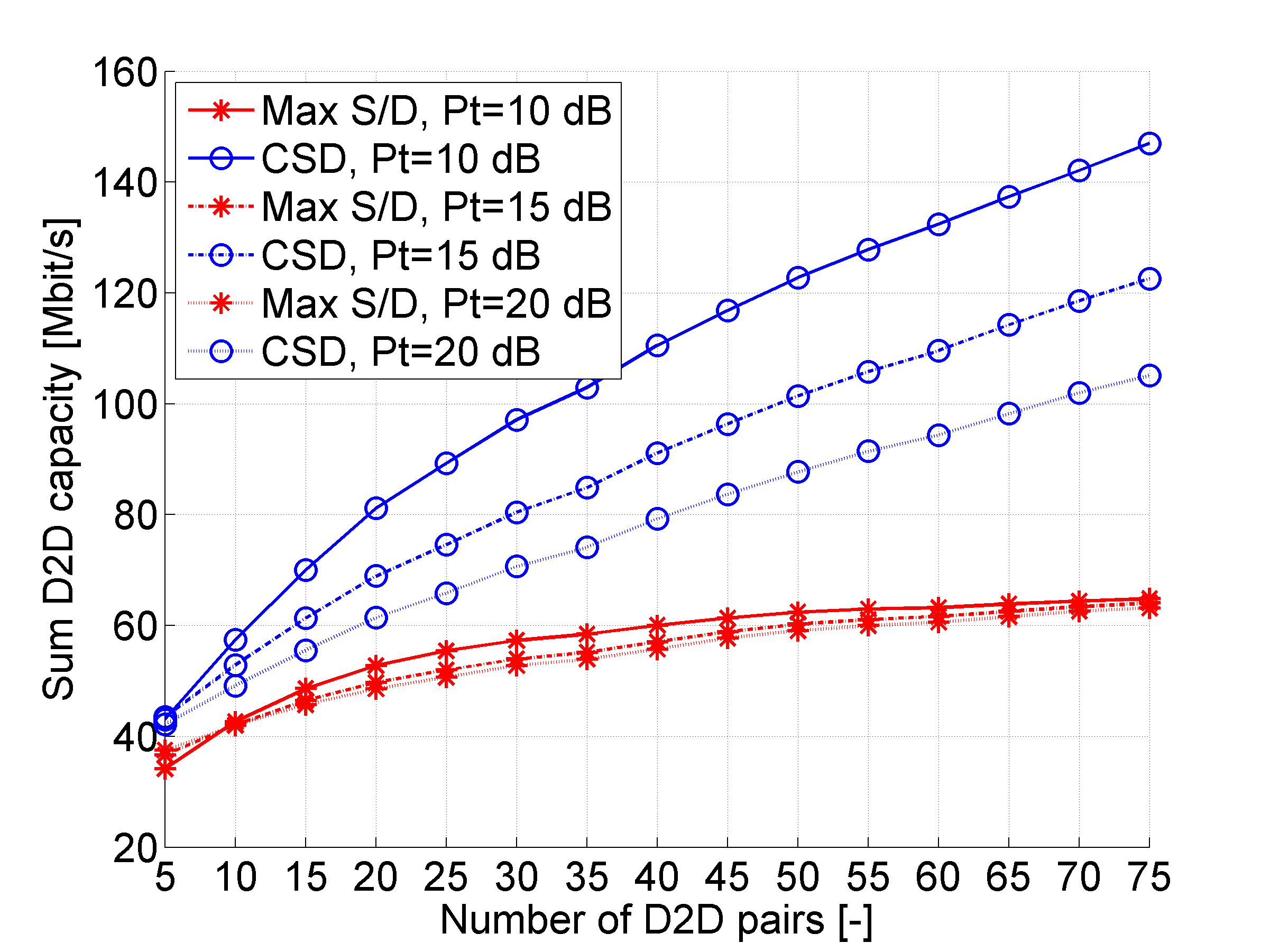}
\caption{Sum capacity depending on number of D2D pairs.}
\label{fig:3}
\end{figure}
\begin{figure*}[t!]
\centering
\begin{subfigure}[t]{0.33\textwidth}
\centering
\includegraphics[width=\textwidth]{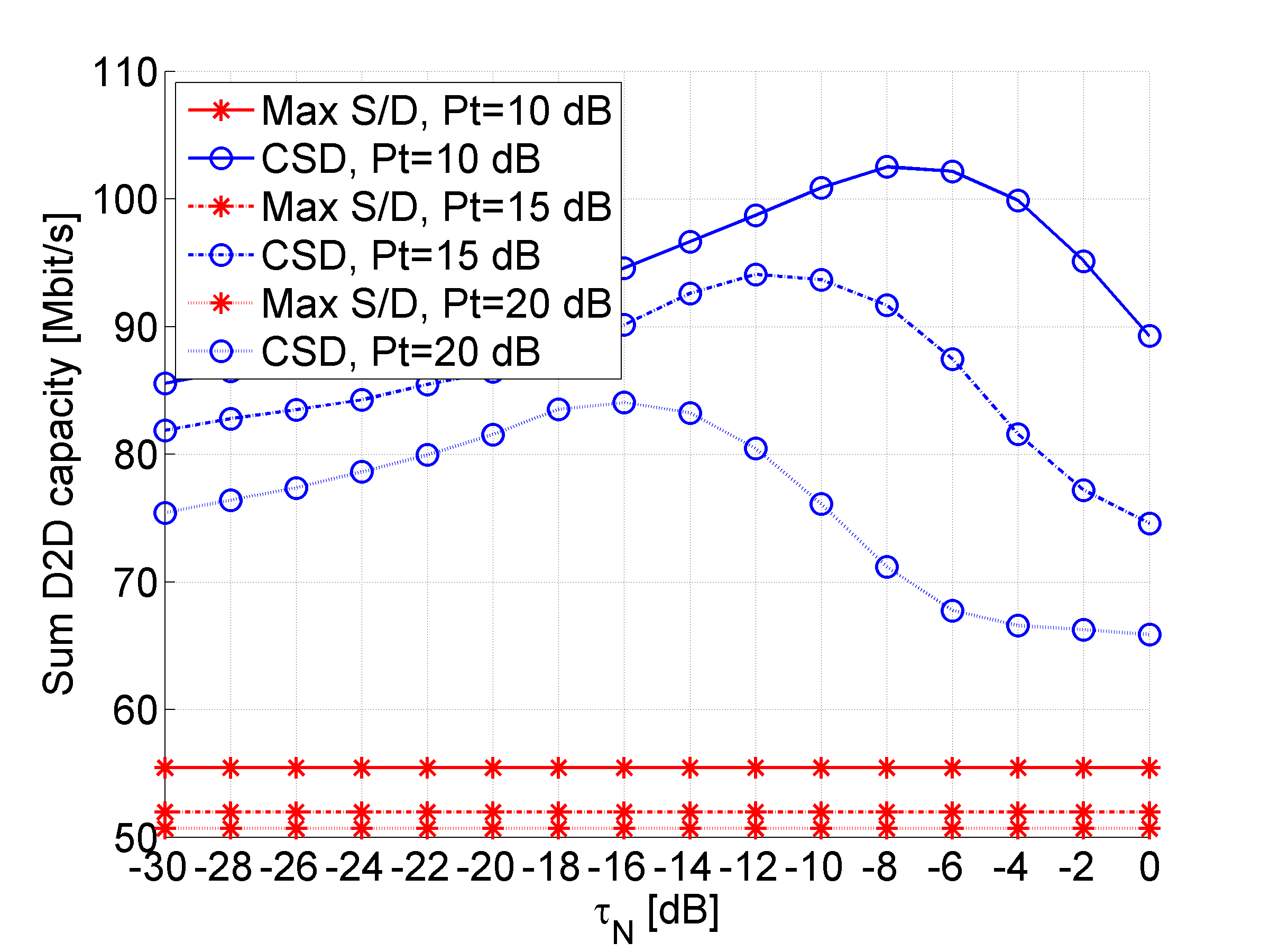}
\caption{25 D2D pairs}
\end{subfigure}\hspace{-1.6em}
~
\begin{subfigure}[t]{0.33\textwidth}
\centering
\includegraphics[width=\textwidth]{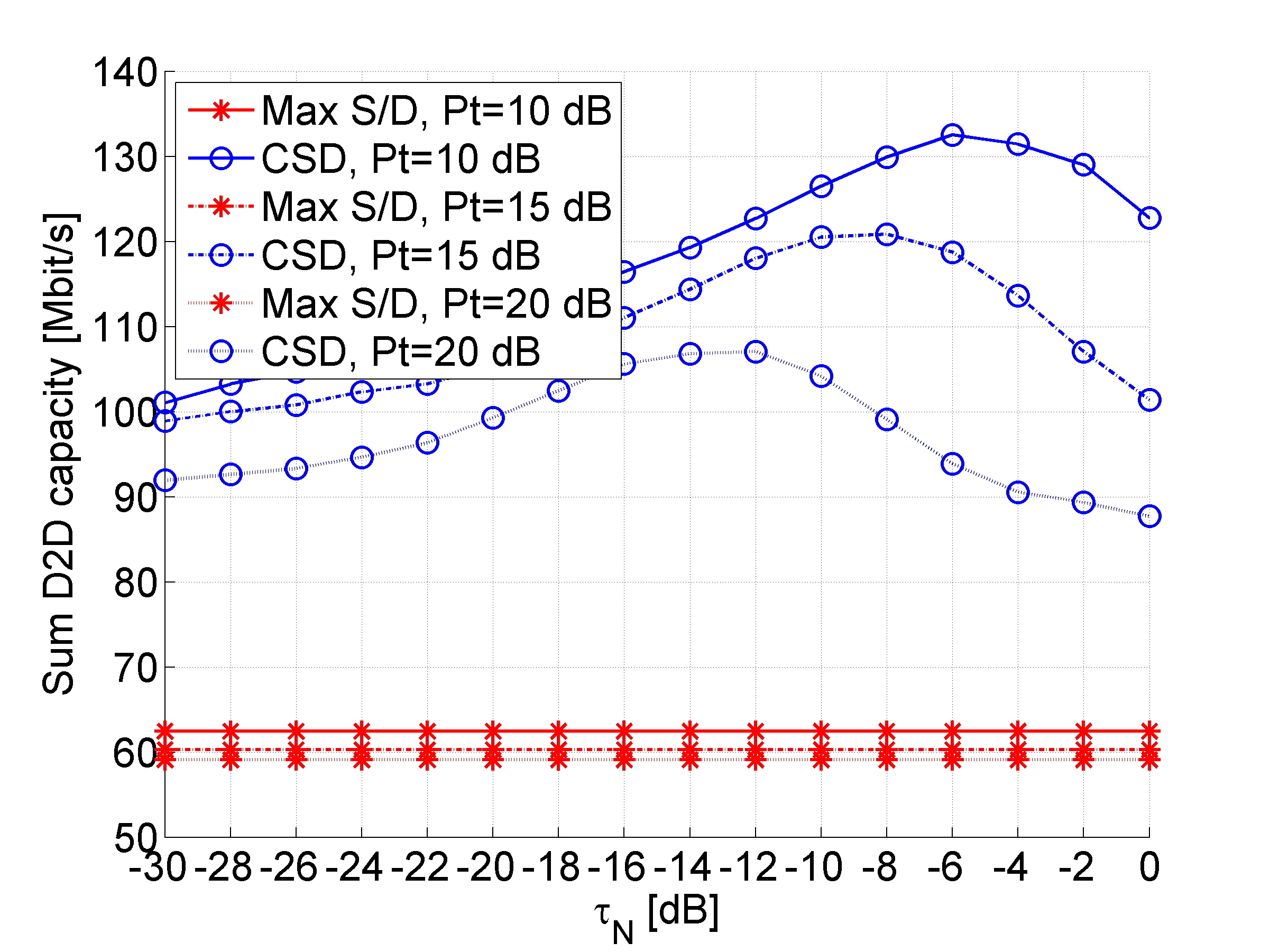}
\caption{50 D2D pairs}
\end{subfigure}\hspace{-0.5em}
~
\begin{subfigure}[t]{0.33\textwidth}
\centering
\includegraphics[width=\textwidth]{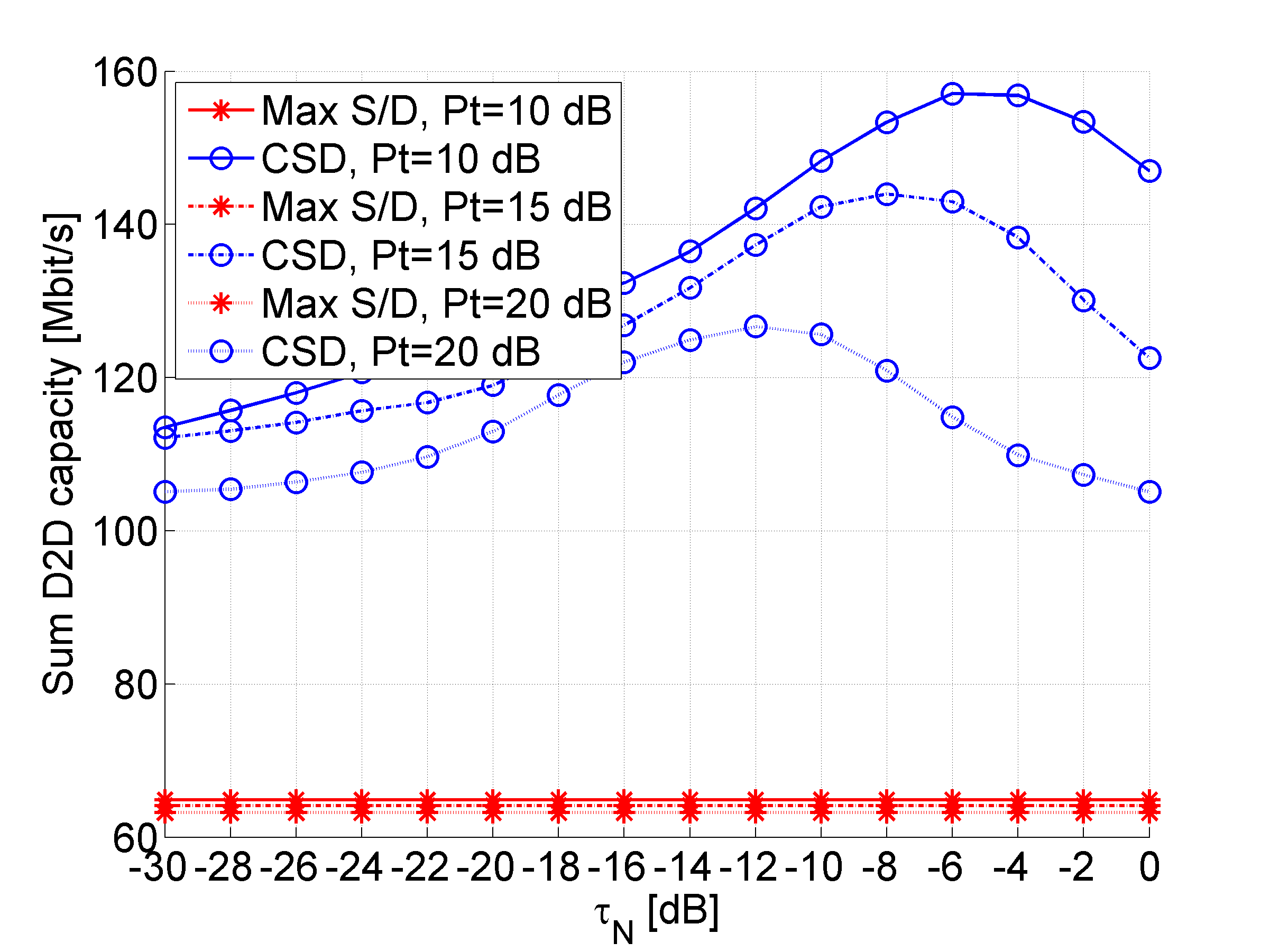}
\caption{75 D2D pairs}
\end{subfigure}\hspace{-0.5em}
\caption{Sum D2D capacity depending on $\tau_N$.}
\label{fig:4}
\end{figure*}
Further, we assume a physical layer data frame with duration of 10 ms and channel bandwidth of 20 MHz like in LTE. Each frame is composed of 2 000 RBs used either for data transmission (75 \% of RBs) or for signaling (25 \% of RBs). We consider that the amount of RBs in the shared and dedicated regions available for the data transmission is split equally, that is, 750 RBs are intended for shared region accessible by the CUEs and the D2D pairs and the other 750 RBs in dedicated region is accessed only by the D2D pairs. 

The calculation of path loss for CUE-eNB, DUE-eNB, CUE-DUE, and DUE-DUE is done according to the models defined by 3GPP for evaluation of the D2D proximity services [18].

The simulation results are averaged out over 200 simulation drops. During each drop, new positions of the CUEs and the D2D pairs are generated. All simulation parameters are summarized in Table I.

\subsection{Simulation results}
The simulations results compare the proposed CSD with schemes based on [9][10], which are the only most recent schemes considering the D2D pairs use solely either dedicated or shared resources so that the capacity is maximized (labeled Max S/D).

Fig.~\ref{fig:3} shows the sum D2D capacity depending on the number of deployed D2D pairs. Notice that we do not analyze the capacity of the CUEs as our objective is to enhance performance of the DUEs while the CUEs are not affected by power restrictions in shared region). The proposed CSD allocation significantly outperforms the existing Max S/D scheme selecting either shared or dedicated resources. If transmission power of the CUEs and the D2D pairs in the dedicated region is set to lower values (e.g., 10 dB in our figure), the CSD brings approximately 2.3 times higher capacity for 75 D2D pairs deployed in the system when compare to Max S/D scheme. Even though the gain of the CSD over Max S/D is lowered if Pt is set to 15 dB (roughly 2 times higher capacity) and 20 dB (roughly 1.67 times higher capacity), the proposed CSD scheme is superior to Max S/D. The main reason why the capacity of the CSD is increasing with a decreasing Pt is that the D2D pairs are more interfered in the shared region by the CUEs and also interference among D2D pairs in the dedicated region is increased as well.

The performance of the CSD is influenced by a setting of the neighborhood threshold $\tau_N$ as already discussed in Section III. While in Fig.~\ref{fig:3} $\tau_N$ = 0 dB is considered, Fig.~\ref{fig:4} shows the impact of $\tau_N$ on the sum D2D capacity. We consider $\tau_N$ varying between -30 dB and 0 dB. Notice that for better clarity, we express in the simulations $\tau_N$ in dB while in (6) $\tau_N$ is without unit . It is seen that the optimal value of $\tau_N$ is depends on the transmission power $Pt$ and also on the amount of D2D pairs in the system (see Fig.~\ref{fig:4}). In general, if the transmission power $Pt$ is increased, $\tau_N$ should be set to lower values to achieve the maximal capacity. For example, if there are 25 D2D pairs in the system, the optimal value for $\tau_N$ is -8 dB, -12 dB, and -16 dB for Pt = 10, 15, and 20 dBm, respectively (see Fig.~\ref{fig:4}a). If the number of D2D pairs is increased to 50 or event to 75, the optimal values of $\tau_N$ are slightly higher (by 2-4 dB) when compared to the case with only 25 D2D pairs in the system. The reason why, $\tau_N$ should be set to lower values for higher $Pt$ is that the D2D pairs are able to reuse resources in shared region more efficiently for lower $\tau_N$ than for higher $\tau_N$.  Moreover, $\tau_N$ should be set to higher values for the higher numbers of D2D pairs because an increased ratio of the resource that can be reused by the D2D pairs (achieved by lower $\tau_N$) is not able to outweigh significant interference among D2D pairs themselves. As in previous figure, the sum D2D capacity is degraded for higher Pt since the D2D pairs are interfered by the CUEs in shared region more significantly.
 
The optimal setting of neighborhood threshold $\tau_N$ also leads to a higher gain achieved by the proposed CSD over the Max S/D scheme. The sum capacity with the optimum threshold $\tau_N$  is increased by 2.5, 2.3, and 2 times for $Pt$ = 10, 15, and 20 dBm, respectively, for 75 D2D pairs. Note that appropriate setting of $\tau_N$ in real network can be considered as future work.  

\section{Conclusions}
This paper has proposed the combined shared/dedicated resource allocation scheme for D2D communications. The proposed CSD scheme allows the D2D pairs to utilize resources in both shared and dedicated regions simultaneously. In addition, the same resources can be exploited by several D2D pairs in order to enhance spectral efficiency of the system and to increase overall sum D2D throughput. In this regard, we have introduced graph theory based framework for the purpose of efficient resource allocation. Within this framework, the eNB creates graph showing neighborhood relations between the CUE and D2D pairs and between individual D2D pairs. After decomposition of the graphs into subgraphs and determination of the maximal cliques, the eNB is able to allocate resources maximizing sum D2D throughput. The results indicate that the D2D capacity can be significantly improved (more than twice) when compared to the scheme selecting only the shared or the dedicated region. 

As a future work, we intend to perform in-depth theoretical analysis for deriving of optimal D2D neighborhood threshold.

\section*{Acknowledgment}
This work has been supported by Grant No. GA17-17538S funded by Czech Science Foundation.

\end{document}